# 64-pixel NbTiN superconducting nanowire single-photon detector array for spatially resolved photon detection

**Shigehito Miki,[1,*] Taro Yamashita,[1] Zhen Wang,[1,2] and Hirotaka Terai[1]**

[1]Advanced ICT Research Institute, National Institute of Information and Communications Technology, 588-2, Iwaoka, Nishi-ku, Kobe, Hyogo 651-2492, Japan

[2]Shanghai Center for SuperConductivity, Shanghai Institute of Microsystem and Information Technology, Chinese Academy of Sciences, R3317, 865 Changning Road, Shanghai 200050, PR China

**Abstract**

We present the characterization of two-dimensionally arranged 64-pixel NbTiN superconducting nanowire single-photon detector (SSPD) array for spatially resolved photon detection. NbTiN films deposited on thermally oxidized Si substrates enabled the high-yield production of high-quality SSPD pixels, and all 64 SSPD pixels showed uniform superconducting characteristics within the small range of 7.19–7.23 K of superconducting transition temperature and 15.8–17.8 μA of superconducting switching current. Furthermore, all of the pixels showed single-photon sensitivity, and 60 of the 64 pixels showed a pulse generation probability higher than 90% after photon absorption. As a result of light irradiation from the single-mode optical fiber at different distances between the fiber tip and the active area, the variations of system detection efficiency (SDE) in each pixel showed reasonable Gaussian distribution to represent the spatial distributions of photon flux intensity.

**1. Introduction**

Recent progress in improving the performance of superconducting nanowire single-photon detectors (SNSPDs or SSPDs [1]) are nothing short of eye-opening. At present, their system detection efficiency (SDE) is reaching unity at a wavelength of 1550 nm with a dark count rate (DCR) lower than 1 kcps [2-4], and a short-system timing jitter of 20–40 ps can be obtained [5]. These values are strikingly superior to those obtained with commonly used avalanche photo diodes (APDs) at telecommunication wavelengths [6]. In fact, practical SSPD systems are already employed in a variety of applications such as quantum key distribution, quantum optics experiments, and so on [7-13]. Meanwhile, arranging the independent multipixels into an active area is an attractive configuration because this allows high-speed operation, enlargement of the device area, pseudo photon number resolution, and spatial resolution. The realization of large-format SSPD arrays will have a great impact on various application fields such as fluorescent imaging, laser ranging, high-resolution depth imaging, free space optical communications, and so on.

One of the critical issues to realizing large-format SSPD arrays is how to reduce heat flow from room temperature through coaxial cables, the number of which increases as of the number of pixels increases in a conventional readout scheme. Therefore, our primary effort so far has been focused on the development of cryogenic readout electronics using a single flux quantum (SFQ) circuit because the required number of cables can be drastically reduced by implementing SFQ circuits and SSPDs in a cryocooler system. We have confirmed correct SFQ operation with output signals from SSPDs [14], and we have succeeded in implementing a four-pixel SSPD array and SFQ circuit in a Gifford–McMahon (GM) cryocooler system with no serious crosstalk [15]. Accordingly, scaling the number of SSPD pixels should be a next step, which will require significant uniformity of superconducting nanowire characteristics. Because our recent development of single-pixel NbTiN SSPDs prepared on a thermally oxidized Si substrate can provide high SDE devices with high yield [4], it is natural to apply this technology to scale up the SSPD arrays. In this work, we report the development of two-dimensionally arranged 64-pixel NbTiN SSPD array and characterization of electrical and optical properties in a GM cryocooler system. We also show the spatial resolution of the 64-pixel SSPD array by irradiating it with incident light from a single-mode optical fiber at different distances between the fiber tip and the active area.

## 2. Experimental procedure

Figure 1 shows the scanning electron micrograph (SEM) of the 64-pixel NbTiN SSPD device. The NbTiN nanowire pixels were fabricated on a Si substrate with a 250-nm-thick thermally oxidized $SiO_2$ layer. Although we chose a thermally oxidized substrate that can configure the double-sided cavity structure by placing a $\lambda/4$ dielectric cavity and mirror on the nanowire [4], they were not embedded in this device in order to exclude the influence of optical absorbance fluctuations that may be caused by each layer's thickness variations or by partial imperfections of the cavity structure. The fabrication process of the NbTiN SSPDs was basically the same as described elsewhere [4,16]. The 5-nm-thick NbTiN nanowire was formed to be 100 nm wide and 100 nm spaced meandering lines covering an area of $5 \times 5$ μm, thus configuring one nanowire pixel. The $8 \times 8$ nanowire pixels were two-dimensionally arranged with spacing of 3.4 μm, covering an area of $63 \times 63$ μm. The 200-nm-wide interconnection lines were formed using the same 5-nm-thick NbTiN nanowires in the spaces between the nanowire pixels, in which the width of interconnection lines was two times wider than that of the nanowire pixel in order to prevent a response to single-photon incidence. The interconnection lines were then connected to coplanar waveguide (CPW) lines. Since the dc resistance of long (~2 mm) CPW lines could lead to a disturbance of the correct operation of the SSPD, 100-nm-thick NbN films with a superconducting transition temperature ($T_c$) of ~15 K were used to assure zero dc resistance.

Figure 2(a) shows a photograph of the chip-mounting block for the 64-pixel SSPD array. As shown in the figure, the 64-pixel SSPD array chip was mounted on the chip-mounting block with a specifically designed printed circuit board (PCB), and each nanowire pixel was wire bonded to a 50 Ω microstrip line on the PCB. Since the total number of coaxial cables introduced into our cryostat system was nine, we characterized the electrical and optical properties of the 64 nanowire pixels by changing the connections between the microstrip lines on the PCB and the coaxial cables in turn. A single-mode optical fiber for a wavelength of 1550 nm was introduced into the cryocooler system, and the end of the fiber was fixed to the rear side of the chip-mounting block after aligning the incident light from the fiber with the device active area, as schematically drawn in Fig. 2(b). We adjusted the distance between the fiber tip and the device active area ($L_{\text{fiber-sspd}}$) to 3 mm and 470 μm in order to observe the incident photon response at the respective distances. The packaged block was cooled with a 0.1 W GM cryocooler system, which can cool the sample stage to 2.3 K [17].

To measure the SDE, a continuous tunable laser was used as the input photon source. The wavelength of the light source was fixed to 1550 nm and attenuated so that the photon flux at the input connector of the cryostat was $10^9$–$10^{11}$ photons/s. Although these values are much higher than usually used for SDE measurements (~$10^6$ photons/s [16]), the incident photon flux to each pixel was low enough to maintain the linearity of the output counts to photon flux due to the low coupling efficiency to each pixel ($P_{\text{couple,pixel}}$). We confirmed the linearity of the output counts to the incident photon flux before deciding on the input photon flux. A fiber polarization controller was inserted in front of the cryocooler's optical input in order to control the polarization properties of the incident photons so as to maximize the SDE. The SDE was determined by the relation SDE = ($R_{\text{output}}$ − $R_{\text{DCR}}$)/$R_{\text{input}}$, where $R_{\text{output}}$ is the SSPD output pulse rate, $R_{\text{DCR}}$ the dark count rate, and $R_{\text{input}}$ the input photon flux rate to the system. The SDE of each pixel was measured individually by changing the connection of the readout components at the outer side of the cryocooler system.

## 3. Results

Figure 3 shows the histograms of the observed $T_c$ [Fig. 3(a)] and the superconducting switching current [$I_{\text{sw}}$, Fig. 3(b)] for the 64 nanowire pixels. As shown in the figure, both $T_c$ and $I_{\text{sw}}$ of all pixels fell within the small ranges of 7.19–7.23 K and 15.8–17.8 μA, respectively. These results indicate that all pixels were fabricated uniformly without significant defects in terms of their electrical properties. We then characterized the optical responses of all pixels. Figure 4(a) shows the SDE as a function of bias current for the 64 nanowire pixels when $L_{\text{fiber-sspd}}$ is adjusted to 3 mm. Although the absolute value of the SDE was low due to $P_{\text{couple,pixel}}$, all pixels showed a response against single-photon irradiation. In addition, bias current dependencies in most of the pixels reaching to plateau, indicating high pulse-generation probability after photon absorption ($P_{\text{pulse}}$). Figure 4(b) shows the histogram of maximum pulse generation probability in each pixel ($P_{\text{pulse,max}}$), the values of which were derived from the fitting using a sigmoid function [4,18]. The sigmoid shape for the bias current dependence of the SDE has been shown empirically, and it is also well fitted to our devices in this work [2,4,19]. As shown in the figure, 60 of the 64 pixels exceeded 90% of approximated $P_{\text{pulse}}$.

The SDE of each nanowire pixel can be expressed as $P_{\text{couple,pixel}} \times P_{\text{abs}} \times P_{\text{pulse}} \times (1 - P_{\text{loss}})$, where $P_{\text{abs}}$ is the optical absorptance into the nanowire and $P_{\text{loss}}$ is the optical loss of the system. Here, the asymptotic system detection efficiency $SDE_{\text{asymp}}$, which is the expected plateau value in bias current dependencies, should only be proportional to $P_{\text{couple,pixel}}$ if $P_{\text{abs}}$ and $P_{\text{loss}}$ are uniform for all pixels, because $P_{\text{pulse}}$ can be treated as 1.0. As described above, the dielectric cavity and mirror layers on the nanowires were intentionally not included in order to retain the variation of $P_{\text{abs}}$ as small as possible. $P_{\text{loss}}$ must be constant over the 64 pixels because we introduced the incident light through one optical fiber to all pixels. Therefore, the spatial distribution of $SDE_{\text{asymp}}$ can clearly reflect the spatial distribution of the incident photon intensity. Figures 5(a) and (b) show the color maps of $SDE_{\text{asymp}}$ over the 64 pixels when $L_{\text{fiber-sspd}}$ is 3 mm and 470 μm, respectively. In the figures, $SDE_{\text{asymp}}$ for each pixel was normalized by the highest value among the 64 pixels. If $L_{\text{fiber-sspd}}$ is 3 mm, the illuminated light from the fiber-end spread and beam waist at the device active area become much larger than the active area of 63 × 63 μm. Since the illuminated light from the single-mode fiber follows the Gaussian beam profile, the active area is exposed to a tiny center area of Gaussian beam where the spatial photon flux intensity distribution can be regarded as approximately flat. On the other hand, illuminated light from the fiber-end at $L_{\text{fiber-sspd}}$ of 470 μm does not spread as compared to that at $L_{\text{fiber-sspd}}$ of 3 mm, and the size of the beam waist at the active area is supposed to be smaller than the size of the active area, resulting in spatial variations of the illuminated light power in the device active area according to the Gaussian distribution. The obtained spatial distribution of $SDE_{\text{asymp}}$ in Figs. 5(a) and (b) clearly represents the spatial photon flux intensity distribution as explained above. Especially, the $SDE_{\text{asymp}}$ distribution in Fig. 5(b) could be well fitted to a Gaussian function by the method of least-squares fit, which is shown in Fig. 5(c), and the beam waist ($2\omega_0$) was estimated to be 19.1 μm at $L_{\text{fiber-sspd}}$ of 470 μm.

Although $SDE_{\text{asymp}}$ was verified to be a good reference for determining the photon flux intensity distribution, a sigmoid function fitting process from the bias current dependency is necessary in order to derive the values, which are unfavorable for realizing a real-time signal processing unit such as an SFQ circuit. For simultaneous 64-pixel operation with real-time signal processing, bias currents supplied to each pixel should be the same in order to employ a simple biasing scheme such as that reported in [20]. Therefore, we next verified whether the actual SDE values at a constant bias current well represent the spatial photon flux intensity distribution. Figures 6(a)–(d) show color maps of the SDE distributions (left) and histograms of $P_{pulse}$ (right) at $L_{fiber-sspd}$ of 470 μm at constant bias currents of 15.0, 15.5, 16.0, and 16.5 μA, respectively. As the bias current increased, the SDEs reached their asymptotic values and the variations over the 64 pixels decreased, enabling to represent the spatial distribution of the photon flux intensity more precisely. However, the number of nanowire pixels that could not operate due to locking into a resistive state ($N_{disabled}$) increased with increasing bias current, resulting in poorer visibility, as shown in the figures. On the other hand, $N_{disabled}$ was zero at the lowest bias current of 15.0 μA. Although the variations of $P_{pulse}$ were larger than those at higher bias currents, the SDE variations still represented the photon flux intensity properly. The value of $2\omega_0$ estimated from Gaussian function fitting was 18.5 μm, which is almost same as that obtained from $SDE_{asymp}$.

## 4. Conclusion

We characterized the electrical and optical properties of a 64-pixel NbTiN SSPD array prepared on a thermally oxidized Si substrate. Our two-dimensionally arranged 64-pixel SSPD array exhibited uniform superconductivity in all pixels and pulse generation higher than 90% in 60 of the 64 pixels. We verified that the spatial distribution of $SDE_{asymp}$ in the 64-pixel SSPD array reasonably represents that of the photon flux intensity by irradiating the light from the fiber tip with different distances to the device active area. In addition, even at a constant bias current of 15.0 μA, we obtained similar SDE distributions as compared to $SDE_{asymp}$ distributions with no disabled pixels. Our next study will be a simultaneous operation of all of 64 nanowire pixels with an SFQ signal processing circuit for real-time spatially resolved photon detection. To accomplish this, the 64-pixel NbTiN SSPD array verified in this work already has favorable features. For example, a bias current of 15.0 μA is available to operate the SFQ circuit [21], and adequate image acquisition of the photon intensity distribution at a constant bias current makes it possible to apply the parallel biasing scheme with few feed lines [18]. Of course, although further improvement of bias current dependencies in the SDE such as those achieved by WSi-SSPDs at a low operation temperature of ~300 mK is more favorable for retaining the variations of $P_{pulse}$ even at low bias current regions [2], the results of this work provide insights into realizing large-format position-sensitive SSPD arrays even at operating temperatures of 2–3 K.

**Acknowledgements**

The authors thank Saburo Imamura and Makoto Soutome of the National Institute of Communications Technology for their technical support.

**References**

1. G. Gol'tsman, O. Okunev, G. Chulkova, A. Lipatov, A. Semenov, K. Smirnov, B. Voronov, A. Dzardanov, C. Williams, and R. Sobolewski, "Picosecond superconducting single-photon optical detector," Appl. Phys. Lett. **79** (6), 705-707 (2001).
2. F. Marsili, V. B. Verma, J. A. Stern, S. Harrington, A. E. Lita, T. Gerrits, I. Vayshenker, B. Baek, M. D. Shaw, R. P. Mirin, and S. W. Nam, "Detecting single infrared photons with 93% system efficiency," Nat. Photonics **7** (3), 210-214 (2013).
3. D. Rosenberg, A. J. Kerman, R. J. Molnar, and E. A. Dauler, "High-speed and high-efficiency superconducting nanowire single photon detector array," Opt. Exp. **21** (2), 1440-1447 (2013).
4. S. Miki, T. Yamashita, H. Terai, and Z. Wang, "High performance fiber-coupled NbTiN superconducting nanowire single photon detectors with Gifford-McMahon cryocooler," Opt. Exp. **21** (8), 10208-10214 (2013).
5. L. You, X. Yang, Y. He, W. Zhang, D. Liu, W. Zhang, L. Zhang, L. Zhang, X. Liu, S. Chen, Z. Wang, and X. Xie, "Jitter analysis of a superconducting nanowire single photon detector," AIP Advances **3**, 072135 (2013).
6. Z. Lu, W. Sun, Q. Zhou1, J. Campbell, X. Jiang and M. A. Itzler, "Improved sinusoidal gating with balanced InGaAs/InP single photon avalanche diodes," Opt. Exp. **21** (14), 16716-16721 (2013).
7. H. Takesue, S. W. Nam, Q. Zhang, R. H. Hadfield, T. Honjo, K. Tamaki, and Y. Yamamoto, "Quantum key distribution over a 40-dB channel loss using superconducting single-photon detectors," Nat. Photonics **1**, 343-348 (2007).
8. M. Sasaki, M. Fujiwara, H. Ishizuka, W. Klaus, K. Wakui, M. Takeoka, S. Miki, T. Yamashita, Z. Wang, A. Tanaka, K. Yoshino, Y. Nambu, S. Takahashi, A. Tajima, A. Tomita, T. Domeki, T. Hasegawa, Y. Sasaki, H. Kobayashi, T. Asai, K. Shimizu, T. Tokura, T. Tsurumaru, M. Matsui, T. Honjo, K. Tamaki, H. Takesue, Y. Tokura, J. F. Dynes, A. R. Dixon, A. W. Sharpe, Z. L. Yuan, A. J. Shields, S. Uchikoga, M. Legre, S. Robyr, P. Trinkler, L. Monat, J.-B. Page, G. Ribordy, A. Poppe, A. Allacher, O. Maurhart, T. Langer, M. Peev, and A. Zeilinger, "Field test of quantum key distribution in the Tokyo QKD Network," Opt. Exp. **19** (11), 10387-10409 (2011).
9. K. Yoshino, M. Fujiwara, A. Tanaka, S. Takahashi, Y. Nambu, A. Tomita, S. Miki, T. Yamashita, Z. Wang, M. Sasaki, and A. Tajima, "High-speed wavelength-division multiplexing quantum key distribution system," Opt. Lett. **37** (2), 223-225 (2012).
10. R. Ikuta, H. Kato, Y. Kusaka, S. Miki, T. Yamashita, H. Terai, M. Fujiwara, T. Yamamoto, M. Koashi, M. Sasaki, Z. Wang, and N. Imoto, "High-fidelity conversion of photonic quantum information to telecommunication wavelength with superconducting single-photon detectors," Phys. Rev. A **87**, 010301(R) (2013).

11. R. Jin, K. Wakui, R. Shimizu, H. Benichi, S. Miki, T. Yamashita, H. Terai, Z. Wang, M. Fujiwara, and M. Sasaki, "Nonclassical interference between independent intrinsically pure single photons at telecommunication wavelength," Phys. Rev. A **87**, 063801(2013).

12. A. McCarthy, N. J. Krichel, N. R. Gemmell, Ximing Ren, M. G. Tanner, S. N. Dorenbos, V. Zwiller, R. H. Hadfield, and G. S. Buller, "Kilometer-range, high resolution depth imaging via 1560 nm wavelength single-photon detection," Opt. Exp. **21** (7), 8904-8915 (2013).

13. N. R. Gemmell, A. McCarthy, B. Liu, M. G. Tanner, S. D. Dorenbos, V. Zwiller, M. S. Patterson, G. S. Buller, B. C. Wilson, and R. H. Hadfield, "Singlet oxygen luminescence detection with a fiber-coupled superconducting nanowire single-photon detector," Opt. Exp. **21** (4), 5005-5013 (2013).

14. S. Miki, H. Terai, T. Yamashita, K. Makise, M. Fujiwara, M. Sasaki, and Z. Wang, "Superconducting single photon detectors integrated with single flux quantum readout circuits in a cryocooler," Appl. Phys. Lett. **99** (11), 111108 (2011).

15. T. Yamashita, S. Miki, H. Terai, K. Makise, and Z. Wang, "Crosstalk-free operation of multi-element SSPD array integrated with SFQ circuit in a 0.1 W GM cryocooler," Opt. Lett. **37** (14), 2982–2984 (2012).

16. Z. Wang, S. Miki, and M. Fujiwara, "Superconducting nanowire single-photon detectors for quantum information and communications," IEEE J. Selected Topics in Quantum Electron. **15** (6), 1741-1747 (2009).

17. S. Miki, M. Fujiwara, M. Sasaki, and Z. Wang, "Development of SNSPD system with Gifford-McMahon cryocooler," IEEE Trans. Appl. Supercond. **19** (3), 332-335 (2009).

18. T. Yamashita, S. MIki, H. Terai, and Z. Wang, "Low-filling-factor superconducting single photon detector with high system detection efficiency," Opt. Exp. **21** (22), 27177-27184 (2013).

19. F. Marsili, F. Bellei, F. Najafi, A. E. Dane, E. A. Dauler, R. J. Molnar, and K. K. Berggren, "Efficient single photon detection from 500 nm to 5 μm wavelength," Nano Lett. **12**, 4799-4804 (2012).

20. T. Yamashita, S. Miki, H. Terai, K. Makise, and Z. Wang, "Parallel bias and readout techniques toward
realization of large-scale SSPD array with SFQ circuit," IEEE Trans. Appl. Supercond. **23** (3), 2500804 (2013).

21. H. Terai, T. Yamashita, S. Miki, K. Makise, and Z. Wang, "Low-jitter single flux quantum signal readout from superconducting single photon detector," Opt. Exp. **20** (18), 20115-20123 (2013).

**Figures**

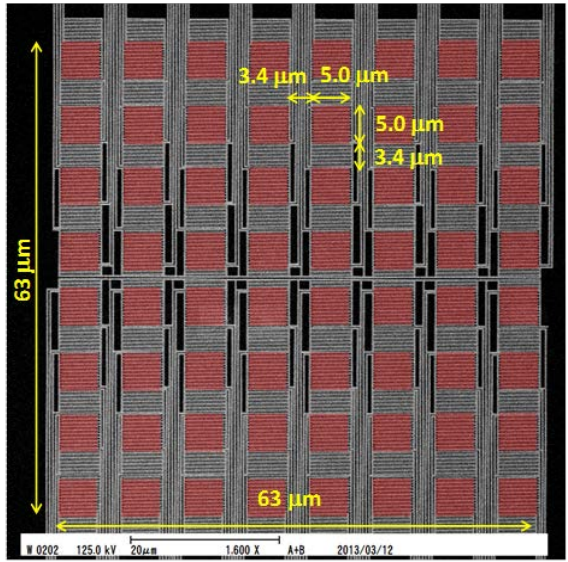

Fig. 1. Scanning electron micrograph of 64-pixel NbTiN SSPD array.

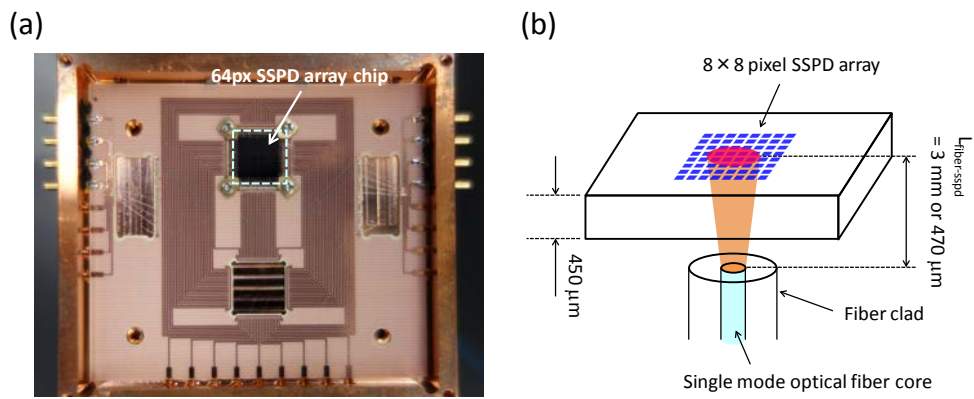

Fig. 2(a) Photograph of chip-mounting block for 64-pixel SSPD array. (b) Schematic drawing of positional relationship between 64-pixel SSPD array, single-mode optical fiber, and illuminated light from the fiber.

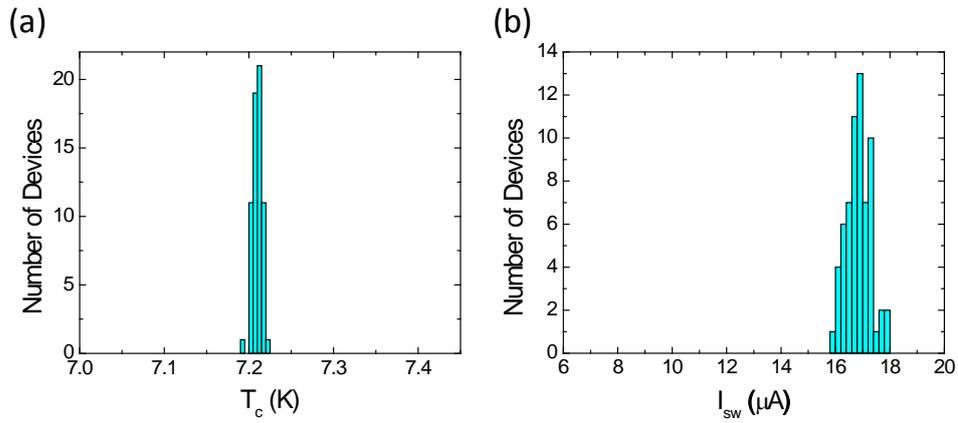

Fig. 3. Histogram of (a) $T_c$ and (b) $I_{sw}$ for the 64 NbTiN nanowire pixels.

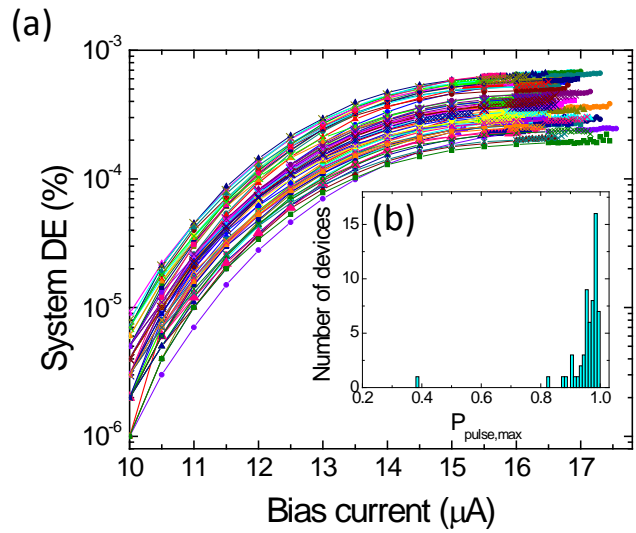

Fig. 4(a) SDE as a function of bias current for the 64 NbTiN nanowire pixels at $L_{\text{fiber-sspd}}$ of 3 mm. (b) Histogram of $P_{\text{pulse,max}}$ for the 64 NbTiN nanowire pixels.

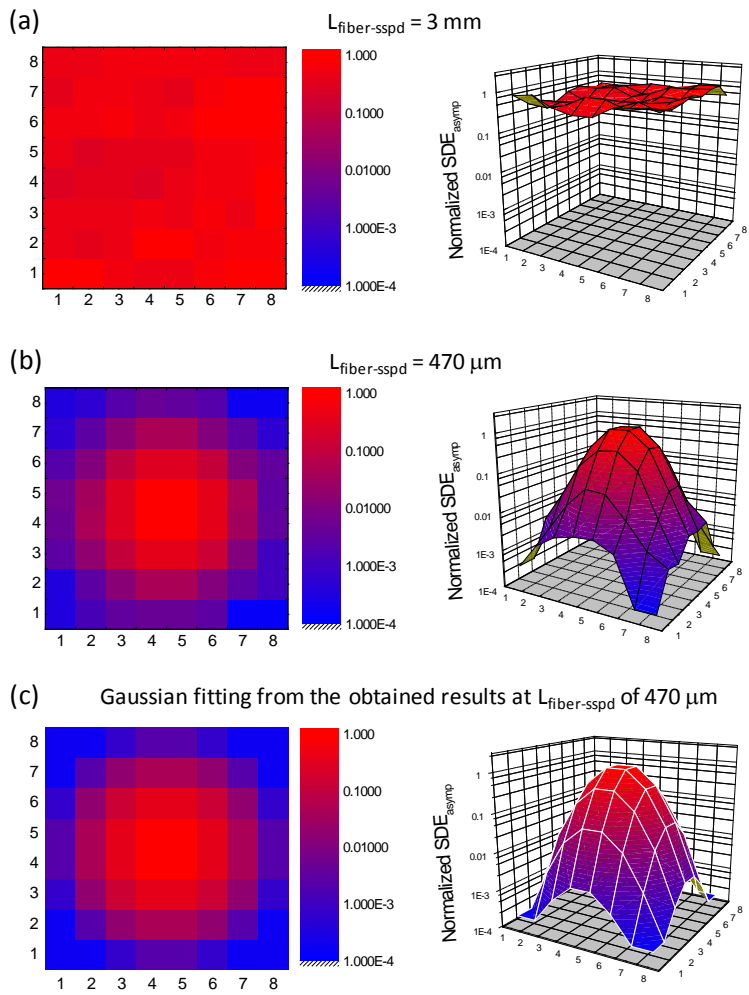

Fig. 5. The color maps of $SDE_{asymp}$ over 64 pixels when $L_{fiber\text{-}sspd}$ is (a) 3 mm and (b) 470 μm. (c) Fitting data of Gaussian function from the results at $L_{fiber\text{-}sspd}$ of 470 μm by means of the method of least squares.

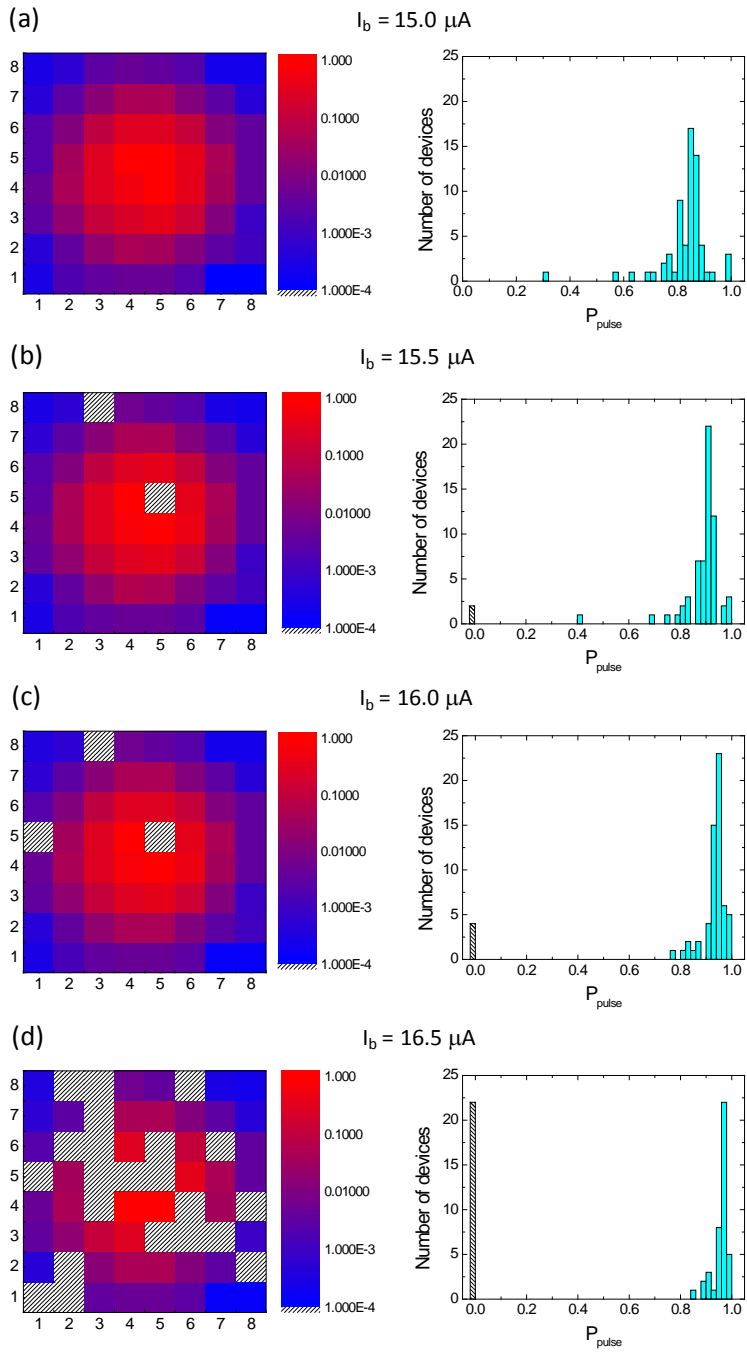

Fig. 6. The color maps of the SDE and histograms of $P_{pulse}$ at $L_{fiber-sspd}$ at 470 μm at constant bias currents ($I_b$) of (a) 15.0, (b) 15.5, (c) 16.0, and (d) 16.5 μA, respectively. The diagonal-lined pixels and bars in the figures indicate the nanowire pixels that could not operate due to locking into a resistive state